# Simultaneous improvements in conversion and properties of molecularly controlled CNT fibres

*Anastasiia Mikhalchan[a], María Vila[a,b], Luis Arévalo[a], Juan J. Vilatela[a]\**

[a] IMDEA Materials Institute, 28906 Getafe, Madrid, Spain
[b] Escuela Técnica Superior de Ingeniería de Telecomunicación (ETSIT), Universidad Rey Juan Carlos, C/Tulipán s/n, 28933 Madrid, Spain

\* Corresponding author. IMDEA Materials Institute, 28906 Getafe, Madrid, Spain
*E-mail address:* juanjose.vilatela@imdea.org (J. J. Vilatela)

## ABSTRACT

Fibres of ultralong and aligned carbon nanotubes (CNT) have axial properties above reference engineering materials, proving to be exceptional materials for application in structural composites, energy storage and other devices. For CNT fibres produced by direct spinning from floating catalyst chemical vapor deposition (FCCVD), a scaled-up method, the challenge is to simultaneously achieve high process conversion and high-performance properties. This work presents a parametric study of the CNT fibre spinning process by establishing the relation between synthesis conditions, molecular composition (i.e. CNT type), fibre mechanical and electrical properties, and conversion. It demonstrates tensile properties (strength 2.1 ± 0.13 N/tex, modulus 107 ± 7 N/tex) above some carbon fibres, combined with carbon conversion about 5%, significantly above literature on similar materials. The combined improvement on conversion and properties is obtained by conducing the reaction at high temperature (1300 °C), using toluene as a carbon source, and through adjustment of the promotor to carbon ratio (S/C) to favor formation of few-layer, collapsed CNTs that maximize packing at relatively high conversions. Lower S/C ratios produce low-defect single-wall CNT, but weaker fibres. Increase in electrical conductivity to 3 x $10^5$ S/m is also observed, with the data suggesting a correlation with longitudinal modulus.

# 1. INTRODUCTION

Assembling CNTs into fibres is the most natural embodiment to exploit their axial properties on a macroscopic scale. Integrating CNTs as fibres and fabrics has also opened new routes for their effective use as components of structural composites [1], batteries [2], sensors [3], etc., that can potentially replace traditional materials.

Since the first development of CNT fibres there has been steady progress in the improvement of their properties through a better understanding of structure-property relations [4]. For fibres spun from liquid crystalline solutions of CNTs, for example, reported axial tensile strength has increased annually by 20 to 25% [5], largely through the development of strategies to spin from solutions of ever longer high-quality CNTs, including those produced industrially by companies such as OCSiAl[1] (pursing commercialization of up to 100 tons of TUBALL nanotubes annually in EU) and Meijo Nano Carbon[2]. Alignment of CNTs is also well established as a dominant factor of axial properties such as tensile modulus [6].

After consistent improvement in CNT fibre properties for nearly two decades the focus of research has broadened to include other aspects of the CNT fibre manufacturing processes, such as reactor engineering and scalability [7]. This is particularly true for the floating catalyst chemical vapour deposition (FCCVD) route, often referred to as the direct spinning method [8]. The interest in the direct spinning method is partly because of increasing number of pilot plants at major industrial facilities, but in addition, because it encompasses the direct transformation of raw precursors into fully-formed high-performance materials in a single fabrication process. As such, it provides complete insight from the input of precursors to the determination of molecular and structural features to the determination of bulk material properties. Thus, current efforts are focused on improving direct spinning as a process, rather than only improving the resulting CNT fibre.

There is a significant body of work that has studied different aspects of the FCCVD reaction under fibre spinning conditions, including the effects of carbon source [4,9–11], role of group 16 elements as promoters [12–14], and the flow rate of carrier gas [15]. For example, higher carbon mole fractions cause excess soot / undesirable products formation, whereas at very low carbon mole fractions the amount of carbon generated upon decomposition of carbon source is insufficient for continuous growth of spinnable CNT aerogel [4,16]. Similarly, control over the predominant type of constituent CNT, in terms of number of layers and diameter, can be achieved by reducing the ratio of promotor to carbon precursor [17], increasing $H_2$ carrier flow rate at constant precursor concentration [15] or adjusting precursor injection [14]. However, most reports target specific fibre compositional features (e.g., $I_D/I_G$ or CNTs type in terms of number of layers), and do not provide a complete picture in terms of synthesis conditions, fibre properties, and very importantly: carbon conversion, namely the fraction of C in the precursor converted into a continuous solid fibre. The data available for FCCVD shows a conversion of carbon in different precursors to CNT fibres ranging from 0.07% to 10% [4]. This is similar to conversions obtained in the early production of highly graphitic carbon black from natural gas (3-6%), but still below present conversion values (10-65%) for related thermochemical reactions [18]. Furthermore, the key is to ensure that high conversions are achieved while retaining or improving the properties of CNT fibres.

---

1 https://ocsial.com/about/
2 https://meijo-nano.com/en/index.php

In this work, we present an extensive process study of the direct spinning FCCVD method by relating synthesis parameters, CNT structural descriptors, bulk mechanical and electrical properties and conversion. The focus is on the effects of reaction temperature, carbon source and on control over CNT type through adjustment of the promotor-to-carbon ratio. We demonstrate bulk mechanical properties superior to many carbon fibres at a conversion (close to 5%) above previous reports, and expose process trends relevant for scale up efforts.

## 2. MATERIALS and METHODS

*2.1 Materials*

Thiophene (extra purity >=99%) used as promoter, and toluene (purity >99%) as carbon source were used as received. Ferrocene (purity 98%) was recrystallized after sublimation and directly dissolved in the precursor solution.

*2.2 Synthesis of CNT fibres*

Carbon nanotube fibres were synthesized in-house by direct FCCVD process in the laboratory-scale reactor with vertical configuration equipped with a mullite tube furnace. This is a single-step gas-phase synthesis method, where the assembly of very long CNTs grow and entangle to form a spinnable aerogel, which is subsequently drawn out of the reactor tube and collected onto a winding bobbin.

In this work, the furnace was heated to two different temperatures: 1250 °C and 1300 °C, called as "low-temperature" and "high-temperature" experiments in the paper. The synthesis conditions were adjusted to investigate varied atomic ratios of sulfur to carbon, while other parameters such as injection rate, hydrogen flow rate, etc. were kept constant. Atomic ratio of Fe/C was very similar (see Supplementary Material, S1 for synthesis parameters), and based on previous results [12] can be considered approximately the same for both precursors.

The CNT aerogel was extracted at the reactor outlet and continuously collected on a rotating bobbin at 28 m/min, and the precise collection time was recorded for all specimens with a stopwatch. The aerogel filaments were then condensed off-line (while still affixed to the bobbin) with isopropyl alcohol and left to dry overnight (Supplementary Material, S2). No additional stretching was applied at the steps of collection or condensation, preserving the intrinsic alignment of CNT bundles in the fibres.

In order to ensure sample reproducibility, the samples CNT fibre material produced during the first hour of spinning were discarded. Control experiments have shown that although composition in terms of CNT type does not change during this time, reaction conversion undergoes steady increase until saturation (Supplementary Material, S3). This is due to the stabilization of precursor injection and the fact that our liquid injection system operates at a very low feed rate of 2.2 mL/hr.

The quantitative characterization and quality control for synthesis optimization are only possible when the spinning process is stable and continuous. Adopting strict internal guidelines on what to consider as reproducible and stable material is a good measure for further process intensification and scale-up. We consider the synthesis process *as stable* when it is possible to collect the visually uniform material for at least 1 hour continuously without aerogel breaking or reaction stop in the gas-phase. In addition, we consider the synthesis *as reproducible*, if it is possible to reproduce the CNT type (e.g., Raman features) across CNT fibres in batches synthesized repeatedly over months, including after replacing the reactor tube (Supplementary Material, S3, Fig. S3.3).

A particular focus of this work is on the comparison of butanol and toluene as C precursors. For CNTF made from toluene, the synthesis conditions studied reflect the space explored to optimize the process for both properties and conversion, by varying reaction temperature and S/C ratio. For CNTF made from butanol, increasing reaction temperature led to unstable spinning. For both types of materials, the drawing rate was close to the optimum in terms of longitudinal properties and ease of fabrication.

*2.3 Characterization of CNT fibres*

All samples were tested as-spun, without any treatments for compaction, purification or functionalization, except the densification of CNT fibre filaments with isopropyl alcohol.

Raman spectroscopy was performed by using Renishaw system equipped with 532 nm laser (2.33 eV); the polarization of the excitation signal was kept parallel to the CNT alignment direction. At least five different locations for each specimen were analyzed and averaged. The spectra were baseline corrected and normalized to the G-band intensity maximum.

SEM images were taken with FIB-FEGSEM dual-beam microscope (Helios NanoLab 600i, FEI) at 5 kV and 0.69 nA. TEM images were taken using a FEG S/TEM microscope (Talos F200X, FEI) operating at 80 kV. Thermogravimetric analysis (TGA) was done with a Q50 (TA Instruments) ramping at 10°C /min from room temperature to 800 °C in air atmosphere (synthetic air).

Tensile properties of individual CNT fibres were characterized by using Favimat tensile testing machine (Textechno, Germany) at 1 min/min, 2 cm gauge length and 0.7 cN/tex pre-tension (1N/tex is numerically equivalent to 1GPa/SG). Prior to tensile tests, linear density of fibres was evaluated vibroscopically with 0.7 cN/tex pre-tension. Additionally, for each batch of different synthesis conditions linear density was verified gravimetrically, for which about 150-300 m of CNT fibres were continuously collected and their mass was measured using a high-precision microbalance.

Electrical resistance of CNT fibres was measured with the two probes technique (as their resistance is in the kΩ range) using a 2450 Keithley source – measurement unit with the distance between contacts of 3 cm. C-solder (Goodfellow) was used to ensure good bonding to the electrical contacts. For every synthesis condition, several samples were measured to ensure reproducibility. Conductivity was calculated by using the linear density of fibres obtained vibroscopically.

## 3. RESULTS

*3.1 Increasing reaction conversion through molecular control*

The starting point of this work is to study the yield of the CNT fibre synthesis process, and more specifically the conversion of the FCCVD reaction, that is, the fraction of C precursor transformed into CNT fibre – expressed as a percentage. This is a convenient metric since the carbon supply is known and the mass of a harvested CNT fibre extracted from the furnace can be determined gravimetrically. The method requires producing around 300 meters of continuous CNT fibre for accurate mass determination using conventional equipment, and evidently, high reproducibility (Supplementary Material, S3). Note that the mass of harvested CNT fibres includes not only graphitic carbon (CNTs) but also carbonaceous impurities and residual catalyst, which are accounted for through thermogravimetric analysis (Supplementary Material, S4) and their presence is indirectly considered when determining bulk fibre properties.

Fig. 1 presents carbon conversion for CNT fibres produced with different S/C atomic ratios. In previous work using 1-butanol as a carbon source and conducting the reaction at 1250°C, we observed an increase in bulk fibre linear density with increasing S/C ratio, directly related to the increasing number of CNT layers through the progression from single-wall carbon nanotubes (SWCNT) to few-layer (8 <) multi-wall carbon nanotubes (MWCNT) [12]. The corresponding conversions are $0.89 \pm 0.30$ % and $4.71 \pm 0.35$ %, respectively. Using toluene as a carbon precursor leads to large improvements in fibre properties (*vide infra,* 135% increase in tensile strength, 269% in fracture energy and 1151% in electrical conductivity, respectively, compared to butanol-spun CNT fibres [19]). It also results in lower conversions for all S/C ratios when the reaction occurs at 1250 °C, but comparable conversion after increasing reaction temperature to 1300 °C and for S/C ratios above 0.0033 S/C ratio.

The synthesis conditions to achieve relatively high conversion (> 4%) from toluene lead to stable spinning conditions. Fig. 1b shows a photograph of a sample of several kilometers in length, before densification. Very importantly, the samples have high fraction of graphitic carbon (>81%wt) with low amount of amorphous carbon and catalyst residue (Supplementary Material, S4), with no evidence of large impurities or other defects found in the CNT fibre samples upon inspection by electron microscopy (Fig. 1c).

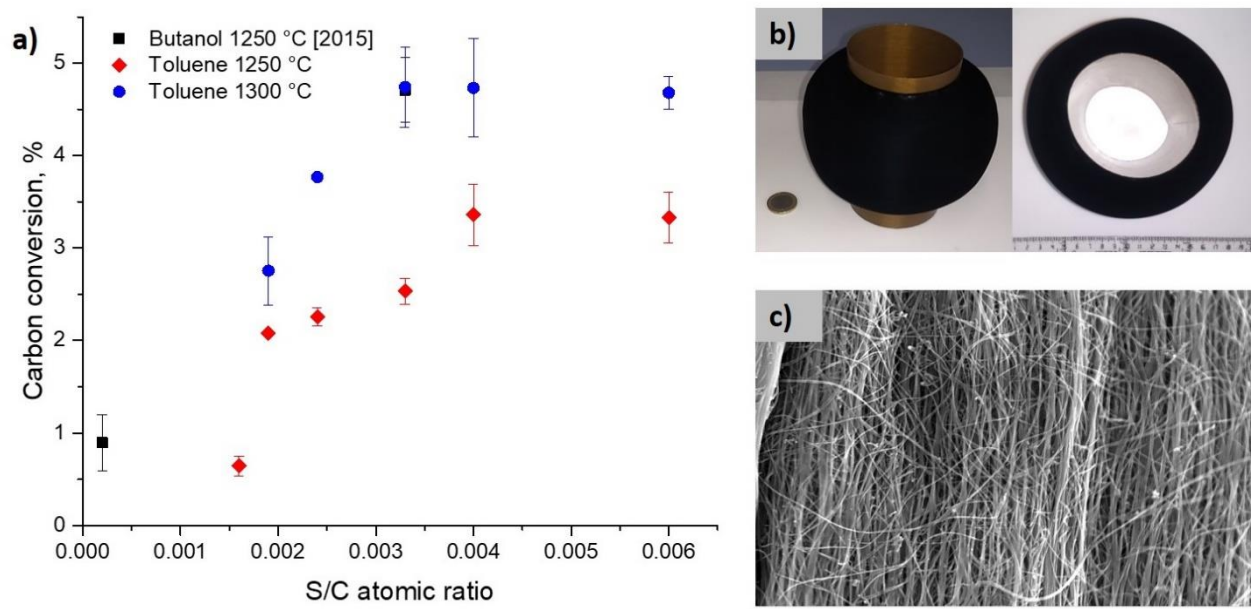

**Fig. 1.** a) Carbon conversion for toluene-based CNT fibres spun at 1250 °C and 1300 °C from the precursors with different S/C atomic ratio and butanol-based fibres [19]; b) a bobbin with ~3 km of CNT filaments continuously spun from toluene during 3 hours; and c) the SEM image of the corresponding densified CNT fibre.

Using Raman spectroscopy, we can relate the trend in conversion to the predominant type of CNT, that is, the increase in conversion through growth of CNTs of more layers. Fig. 2 shows the normalized Raman spectra of CNT fibres produced from toluene at different synthesis temperatures from precursors with different S/C atomic ratios. At lower sulfur content in the precursor mixture the spectra are characteristic of SWCNT, with the presence of radial breathing modes (RBMs) and the superposition of $G^+$ and $G^-$ components of the G band from both metallic and semiconducting SWCNT. The RBMs observed under 532 nm irradiation correspond to diameters of 0.9 to 1.7 nm, although the actual diameter distribution is significantly broader [20]. With higher S/C ratios (0.0033 to 0.0060), the characteristic RBM signals almost disappear, while the G band loses resolved features from longitudinal and tangential modes and instead shows a hint of a $D^*$ shoulder. This trend is also observed with 785 nm laser line (Supplementary Material, S5). Note that all the spectra have low $I_D/I_G$ ratio at the level of 0.1-0.2, implying a high degree of perfection of the graphitic layers.

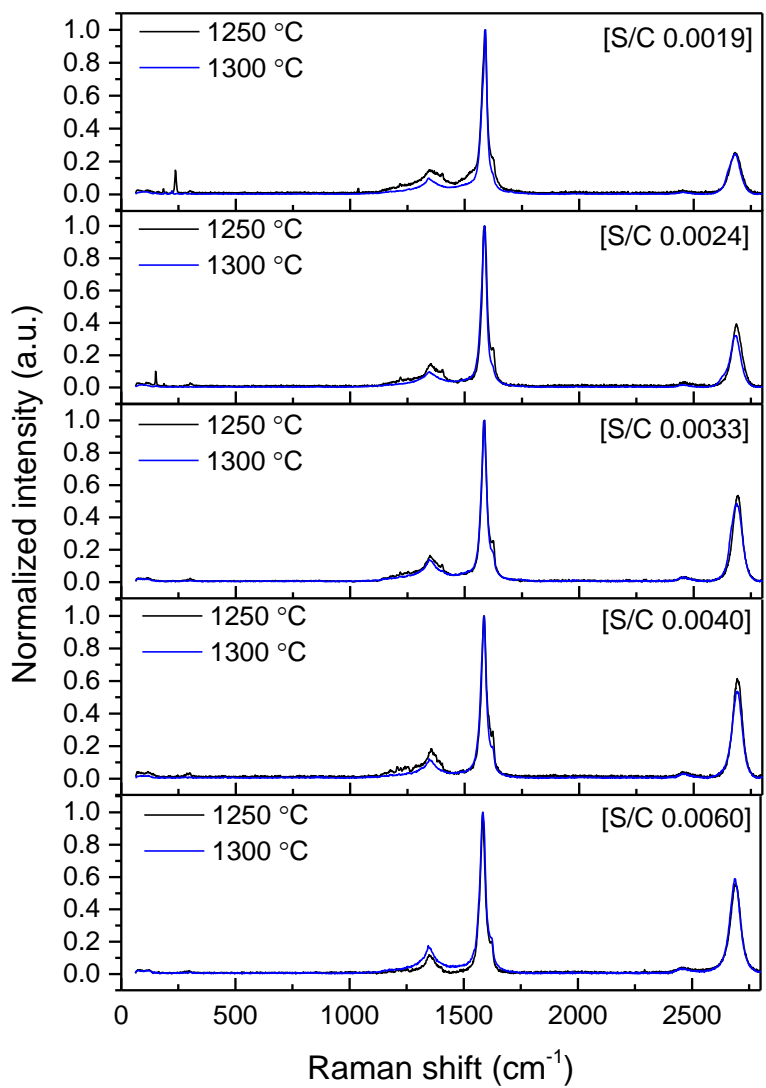

**Fig. 2.** Normalized Raman spectra of CNT fibres synthesized from toluene at 1250 °C and 1300 °C showing the evolution from SWCNT to MWCNT with increasing S/C ratio in the precursor.

The gradual transition from SW to few-layer carbon nanotubes can be conveniently traced through the positions of either the G or 2D Raman peaks. As shown in Fig. 3, there is a clear shift in the position of these two bands for different S/C ratios. There is a notable drop in G band position with a corresponding upshift of the 2D band with increasing S/C ratios. The 2D upshift is related to the increasing number of CNT layers, confirming a change from SWCNT to MWCNT with increasing sulfur concentration, with the similar trends observed for both synthesis temperatures. The data obtained for the recently produced toluene materials in Fig. 3 were confirmed by numerous measurements across different batches and synthesis days. The data correlate well with the data points for 1-butanol-spun fibres from previous work [12]. These results confirm the role of promotor in controlling the predominant type of CNT, as observed using different carbon precursors (methane [11,13,14], toluene [13], 1-butanol [12,19], acetone [14], benzyl alcohol [15]) regardless of the synthesis temperature. Effectively, a small range of values of S/C ratio give access to the whole envelop of CNT types in terms of number of layers (see Supplementary Material, S6 for a TEM example of a distribution of CNT layers and its effect on bulk fibre liner density), although the exact mechanism is not fully understood. Note, however, that there is an apparent "saturation" (here at S/C = 0.004) in the Raman spectra, which does not reflect the actual additional increases in number of layers after further increases in S/C ratio. For a large number of layers the 2D peak position is expected to converge to values similar to those of graphite, thus eventually losing its use as an indicator of the average morphology of the CNT distribution.

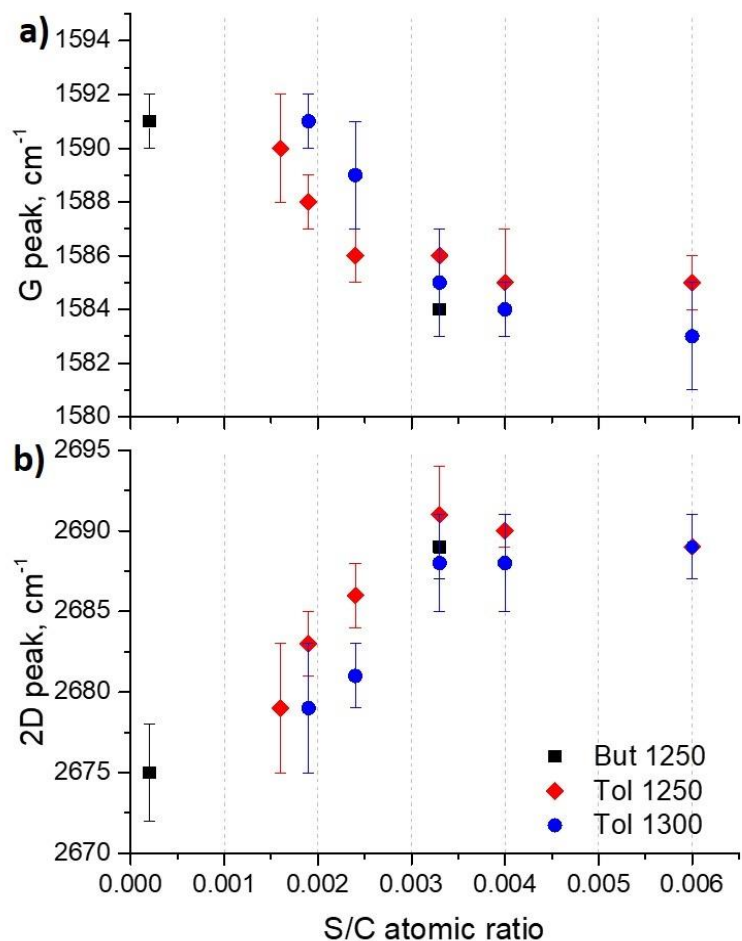

**Fig. 3.** Plot of Raman G- peak (a) and 2D-peak position (b) for samples produced with different amounts of sulfur promotor and at different synthesis temperature.

For both synthesis temperatures, the observed shift in 2D position correlates very well with carbon conversion (Fig. 4). At low-to-medium sulfur/carbon ratios, the carbon conversion gradually increases with increasing number of CNT layers until the point with predominant composition of MWCNTs. Very importantly, this representation enables separation of two effects contributing to conversion: number of CNT layers and increase in CNT number concentration/length. Increases in conversion at constant 2D peak position imply that more CNTs of the same type are produced in the reaction, for example through a larger number of catalyst particles being active. On the other hand, increased conversion through synthesis of CNTs with more layers, for example by changing the composition/size of the catalyst particles but keeping their number concentration constant, would follow a nearly linear dependence on CNT number of layers and diameter. The observed trend of increased conversion with increasing 2D position (increasing S/C ratio) is thus a reflection of the addition of layers to the CNTs. In contrast, the higher conversion observed at 1300 °C compared to 1250 °C for all S/C ratios indicates that either more CNTs are formed or that they are longer as a consequence of the higher reaction temperature.

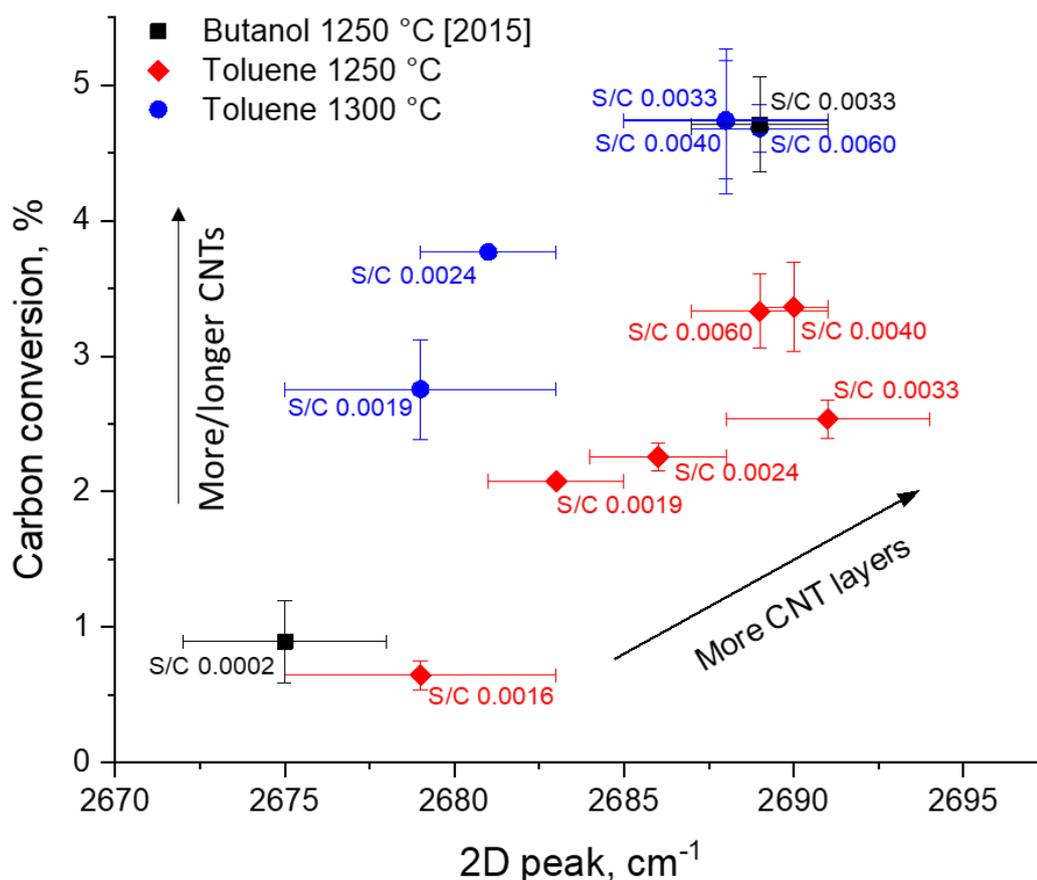

**Fig. 4.** Correlation between carbon conversion and 2D-peak Raman shift for toluene-based CNT fibres produced at 1250 °C and 1300 °C at different S/C ratios compared to butanol-spun fibres from [19].

*3.2 Effect of composition on mechanical properties of CNT fibres*

The interest then is in relating these synthesis conditions to longitudinal CNT fibre properties. The plots in Fig. 5 show values of specific strength, specific modulus, fracture energy and electrical conductivity for the set of CNT fibres synthesized at two different temperatures and at different S/C ratio, as well as previous data using a different C precursor. For each synthesis condition, a few batches from different synthesis days and at least 5-6 fibres from each batch were tested (histograms included in Supplementary Material, S7). Overall, the graphs nicely show the transition from low-strength and low-modulus fibres spun of predominantly SWCNT to high-strength, high-modulus and high toughness at higher S/C ratios. An important conclusion is that there is no intrinsic advantage of manufacturing a fibre of predominantly single-wall CNTs by FCCVD, at least, with respect to mechanical performance. Fibres of predominantly SWCNT spun from butanol have the average specific tensile strength of only 0.18 N/tex, modulus of 12 N/tex and fracture energy of 8.4 J/g; the toluene-based fibres spun at 0.0016 S/C exhibited 0.39 N/tex strength, 17 N/tex modulus, and 11 J/g fracture energy on average. These mechanical properties are similar to fibres comprised of SWCNTs with a diameter <2 nm (1.2 – 1.6 nm) produced by FCCVD in a different reactor at higher $H_2$ flow rates, with resultant specific tensile strength of around 0.15 N/tex [15,21]. The mechanical properties observed for predominantly SWCNT-fibres are significantly below those achieved for the fibres spun from toluene at 1300 °C with mixed population of few-layer CNTs. At 0.0033 S/C ratio, their average specific strength and tensile modulus reached 2.1 N/tex (maximum observed 2.38 N/tex) and 107 N/tex (maximum observed 117 N/tex), respectively, with the fracture energy being in the range of 50-80 J/g. These results confirm that bulk properties of the CNT fibres are dominated by inter-particle rather than intra-particle features, and that degree of CNT perfection, for example as given by the intensity of D/G ratios, is not an adequate predictor of tensile properties. CNT fibres are fibrillary solid with mechanical failure through pull out of CNTs and no evidence of CNT fracture [6,22,23]; hence, the defect density of constituent CNTs does not directly limit tensile strength.

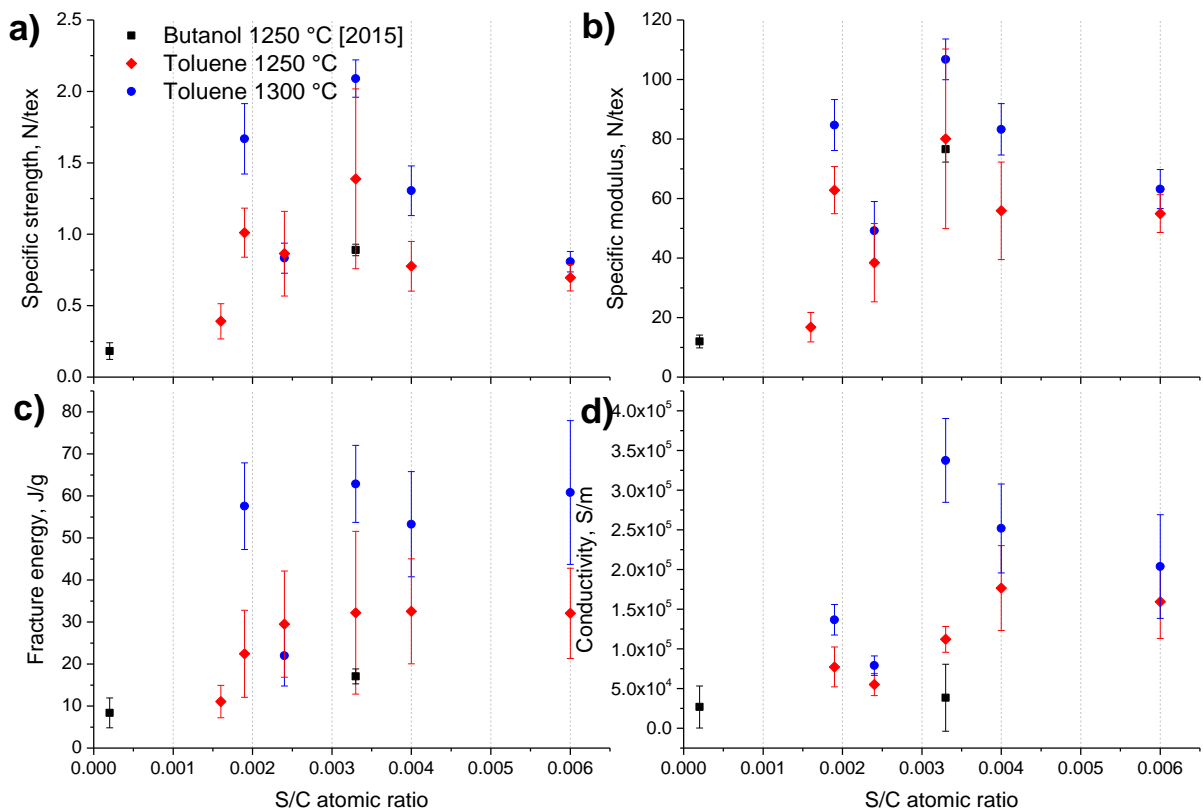

**Fig. 5.** Plots of longitudinal CNT fibre properties of different CNT type, produced through adjustment of the S/C ratio at the point of synthesis: a) specific tensile strength, b) specific modulus, c) fracture energy and d) electrical conductivity for toluene-based CNT fibres produced at 1250 °C and 1300 °C and examples of butanol-spun fibres from [19].

We also note the significant increase in all longitudinal properties for higher synthesis temperatures and when using toluene instead of butanol as a carbon source. Since the comparison is now made across different types of CNTs and at equivalent degree of alignment (note, all fibres were collected at the same constant drawing rate of 28 m/min, and not stretched/post-processed to increase alignment), we attribute this improvement in properties to improved CNT packing and/or longer CNT length. Indeed, few-layer CNTs that collapse into ribbons [24] show more efficient packing [22,25] and thus form longer domains for stress transfer, which would translate into simultaneous increases in tensile strength and modulus, as observed here (Supplementary Material, Fig. S7.2). On the other hand, it remains a challenge to determine length of individual constituent CNTs produced by FCCVD, particularly when assembled as dense fibres. The data also show a general correlation between tensile modulus and longitudinal electrical conductivity (Fig. 6), which may also be attributed to improved packing and longer CNTs reducing internal electrical resistance through elimination of junctions.

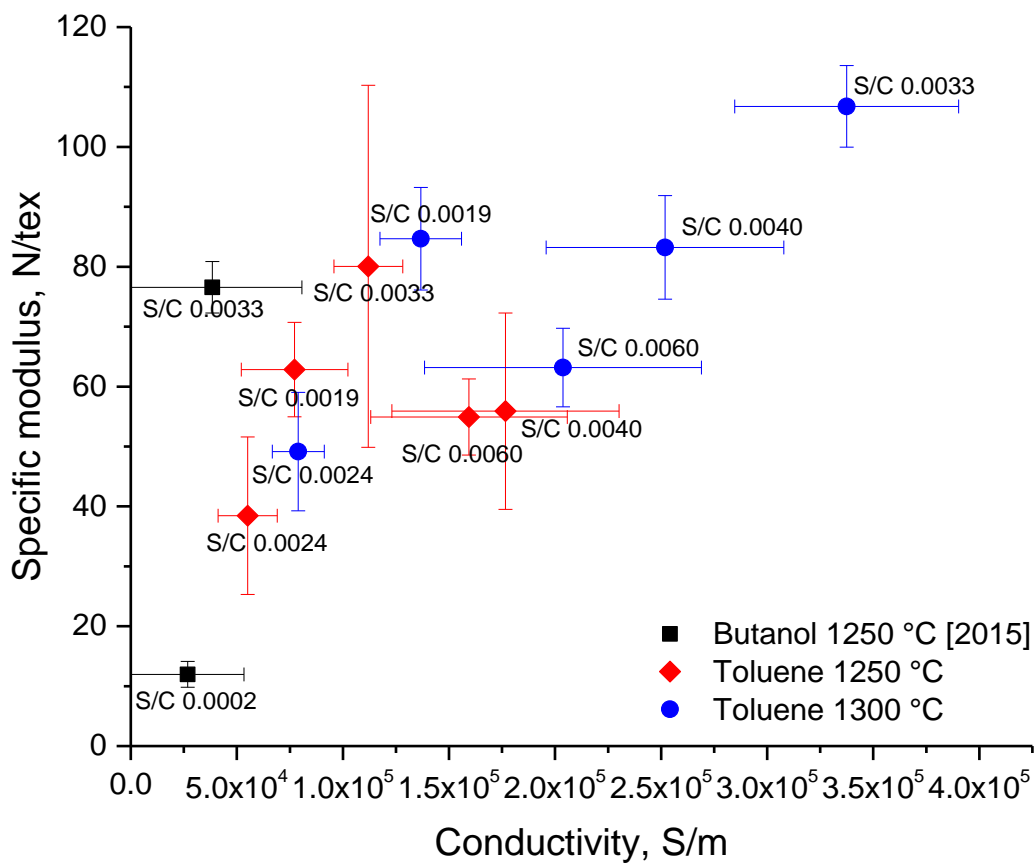

**Fig. 6.** Plot of specific tensile modulus and longitudinal electrical conductivity for the different samples in this study (toluene-based CNT fibres produced at 1250 °C and 1300 °C and examples of butanol-spun fibres from [19]).

*3.3 Analysis in multi-parameter space and comparison to other materials and processes*

The combined data obtained enable establishing the relation between synthesis-structure-properties-conversion. There are various possible comparisons depending on the parameters of interest (data in Supplementary Material, S8), but the overall result is that the synthesis conditions reported here produce a generalized improvement in all metrics. As example, in Fig. 7 we present plots for specific strength and electrical conductivity against conversion. The highest conversion achieved (close to 5%) corresponds to fibres with significantly higher mechanical properties and conductivity than our previous reports [26].

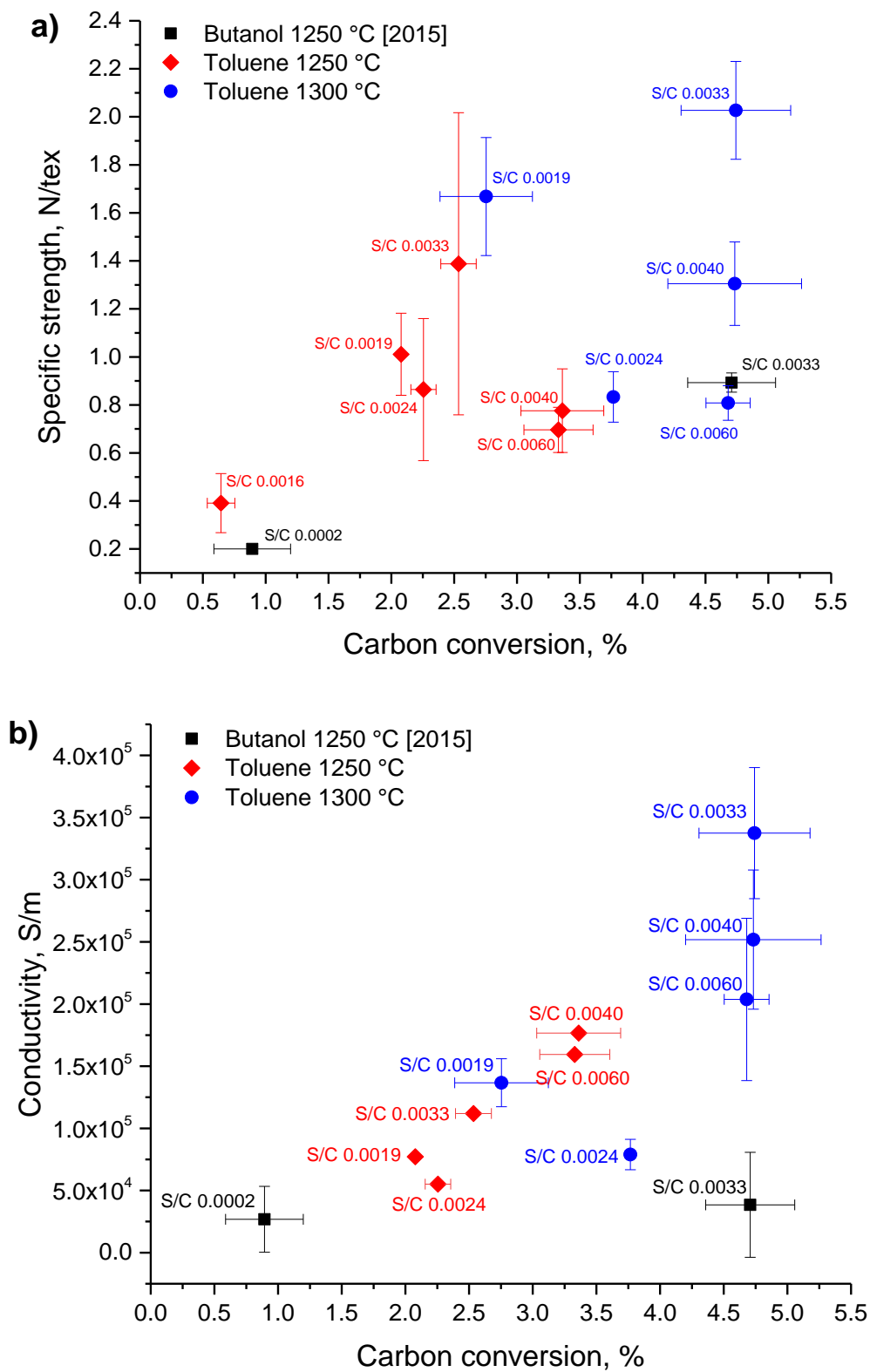

**Fig. 7**. Plots of properties for CNT fibres produced with different molecular composition and under conditions leading to different carbon conversion. a) Specific strength and b) longitudinal electrical conductivity against carbon conversion. Note the general improvements relative to previous synthesis conditions [19].

An interesting feature of CNT fibres is their resemblance to macromolecular solids. Their internal structure and micromechanical behavior are in many respects more similar to that of high-performance polymer fibres than carbon fibres (CF). An implication is that CNT fibres show extremely high fracture energy, coupled with high modulus. Recently, S.H. Lee et al. [33] have demonstrated production of CNTs by FCCVD in the range 1 – 6 mg/min by operating at higher feedstock rate using methane as carbon source. However, the FCCVD process was only used to synthesize CNTs, which were later heat-treated and purified to remove amorphous carbon and iron catalyst particles, and then spun from liquid crystalline solution in chlorosulfonic acid, resulting in CNT fibres with high elastic modulus (230 N/tex) and high fracture energy. The comparison in Fig. 8 shows that present properties of laboratory-scale CNT fibres are above reference synthetic fibres, such as polyaramid, and on the level of ultrahigh-molecular weight polyethylene fibres. This comparison gives testimony of the enormous improvement in bulk fibre properties. Note, in addition, that the samples in the present study are tested as-made, without using post-spin stretching methods [27,28], which may provide further improvement in longitudinal properties.

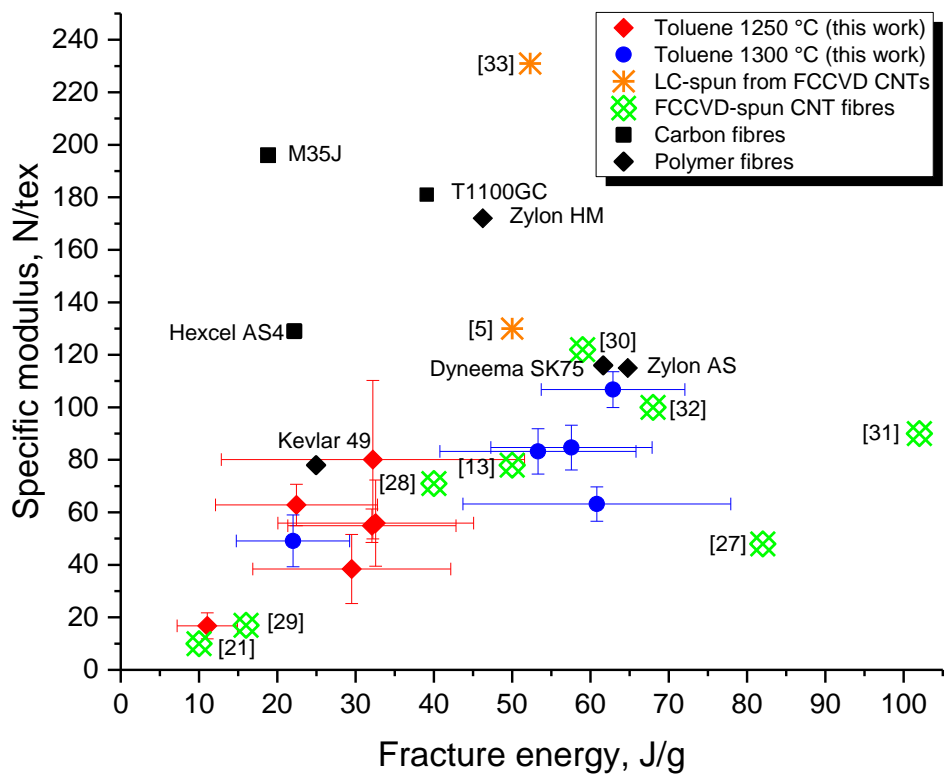

**Fig. 8**. Comparison of tensile specific modulus and tensile fracture energy for CNT fibres and related high-performance polymer fibres and CF. Literature data are shown for the pristine CNT fibres spun by FCCVD [13,21,27–32] and liquid-crystalline fibres spun from CNTs produced by FCCVD [5,33].

Finally, it is of interest to compare current values of yield and carbon conversion in this work with reference thermochemical processes. As shown in Table 1, the FCCVD method has similar performance to manufacturing CF from lignin or rayon, albeit producing much stronger fibres. However, the process yield is still far from optimized thermal process for making CB (with very high methane conversion [34]) or CF processes.

**Table 1.** Process yield, carbon conversion, and tensile properties for conventional carbon fibres and FCCVD CNT fibres

| Fibre | Typical process yield*, % | Carbon Conversion*, % | Exemplary Tensile strength, N/tex | Exemplary Tensile Modulus, N/tex | Ref. |
|---|---|---|---|---|---|
| Carbon PAN | 50 - 60 | 74 - 81 | 3.9 | 180 | [35]; T1100G [36] |
| Carbon MPP | ~85 | ~92 | ~1.8 | ~500 | [35]; K1100 [37] |
| Carbon rayon | 10 - 30 | 22 - 68 | ~0.3 – 0.8 | ~60 | [35]; [38] |
| Carbon kraft lignin | 3 – 15 (up to 45) | 4 – 23 (up to 69) | ~ 0.2 – 0.3 (up to 0.6) | 17 – 35 (up to 45) | [39,40] / [41] |
| CNT FCCVD (Butanol) | 3.05 | 4.71 | 0.89 | 76 | [19] |
| CNT FCCVD (Toluene) | 4.28 | 4.74 | 2.1 | 107 | This work |
| Carbon black (natural gas)§ | 3 - 6 | | N/A | | [18] |
| Carbon black (oil) | 10 – 35# to 65† | | N/A | | [18] |
| Carbon black (natural gas) | 35 - 60 | | | | [34] |

*Notes: \*Here: the process yield for commercial carbon fibres is yield of fibre after all high-temperature treatment(s) of polymeric precursor, The exact composition of each precursor and technological steps vary from one company to another and are generally considered a trade secret, therefore, the table represents only typical ranges of process yield and estimated carbon conversion for different precursors, for reference only. § Here: Historical example of the oldest channel black process being used in the USA since the late 1800s based on the incomplete combustion of natural gas, in contrast to modern optimized oil furnace and thermal processes. # for high quality high surface area pigment CB; † for typical reinforcing CB.*

**CONCLUSIONS**

This work introduces an extensive study towards the improvement of the direct spinning process to make continuous, macroscopic fibres of CNT by drawing them directly from the gas phase during growth by floating catalyst chemical vapor deposition. Using as a baseline a highly-reproducibly method to spin uniform fibres on the kilometer range, we analyze the strategies to control the FCCVD reaction in order to tailor the composition of the fibres in terms of CNT layers and simultaneously improve bulk properties and carbon conversion. We compare experimental data gathered over years in the improvement of synthesis and spinning of CNTF from different precursors: recent data produced in the optimization for toluene, and earlier work using butanol.

The study analyses fibres produced from these precursors, synthesized and spun under comparable conditions, although they need not represent the same level of optimization.

The results demonstrate the predominant role of the promotor/carbon ratio to control CNT number of layers, as observed through the Raman spectra and condensed in the G or 2D band positions. This effect is observed for different precursors (butanol and toluene) and for different reaction temperatures. Equipped with this compositional probe, we could establish that most increases in conversion observed here are due to increasing number of layers in the constituent CNTs.

Replacing butanol with toluene as carbon source produced a large improvement in tensile properties and longitudinal electrical conductivity, with conversion as high as 5%. The resulting mechanical properties are in the range or above other high-performance fibres, particularly in terms of tensile fracture energy. Conversion remain low, in the range of early processes for CB synthesis or fabrication of CF, for example from lignin or rayon, although with tensile strength already at the low end of commercial CF.

**CRediT AUTHORSHIP CONTRIBUTION STATEMENT**

Anastasiia Mikhalchan: Conceptualization, Methodology, Investigation, Data curation, Writing – original draft, Writing – review & editing. María Vila: Methodology, Investigation. Luis Arévalo: Methodology, Investigation, Data processing. Juan José Vilatela: Supervision, Conceptualization, Writing – original draft, Writing – review & editing, Project administration, Funding acquisition.

**DECLARATION OF COMPETING INTEREST**

The authors declare that they have no known competing financial interests or personal relationships that could have appeared to influence the work reported in this paper.

**ACKNOWLEDGMENTS**
A.M. acknowledges funding from the European Union's Horizon 2020 research and innovation program under the Marie Skłodowska-Curie grant agreement 797176 (ENERYARN). M.V. acknowledges the Madrid Regional Government (program "Atracción de Talento Investigador", 2017-T2/IND-5568) for financial support. J.J.V. is grateful for generous financial support provided by the European Union Horizon 2020 Program under grant agreements 678565 (ERC-STEM) and Clean Sky-II 738085 (SORCERER JTI-CS2- 2016-CFP03-LPA-02-11), by the MINECO (RyC-2014-15115, HYNANOSC RTI2018-099504-A-C22), Spain and FOTOART-CM P2018/NMT-4367, Madrid Regional Government.

# Appendix A. Supplementary data

Supplementary data to this article can be found online at https://doi.org/10.1016/j.carbon.2021.04.033